\documentclass[12pt,showpacs,prd]{revtex4}
\usepackage{epsfig,amssymb,amsfonts}

\begin{document}
\title{Coupled-channel model for charmonium levels\\
and an option for $X(3872)$}
\author{Yu. S. Kalashnikova}
\affiliation{Institute of Theoretical and Experimental Physics, 117218,\\
B.Cheremushkinskaya 25, Moscow, Russia}
\newcommand{\be}{\begin{equation}}
\newcommand{\bea}{\begin{eqnarray}}
\newcommand{\ee}{\end{equation}}
\newcommand{\eea}{\end{eqnarray}}
\newcommand{\vep}{\mbox{\boldmath${\rm p}$}}
\newcommand{\veL}{\mbox{\boldmath${\rm L}$}}
\newcommand{\veS}{\mbox{\boldmath${\rm S}$}}
\newcommand{\vek}{\mbox{\boldmath${\rm k}$}}
\newcommand{\ds}{\displaystyle}

\begin{abstract}
The effects of charmed meson loops on the spectrum of charmonium are
considered, with special attention paid to the levels above open-charm
threshold. It is found that the coupling to charmed mesons generates a
structure at the $D \bar D^*$ threshold in the $1^{++}$ partial wave. The
implications for the nature of the $X(3872)$ state are discussed.
\end{abstract}
\maketitle

\section{Introduction}

The charmonium spectroscopy has again become a very interesting field. On
one hand, the $2^1S_0$ $\eta_c'$ state was discovered by Belle
\cite{Belle} and confirmed by BABAR \cite{Babar} and CLEO \cite{Cleo1},
and the $1^1P_1$ $h_c$ was observed in Fermilab \cite{FNAL} and
by CLEO \cite{Cleo2}. The masses of these long missing states are
in perfect agreement with the predictions of quark model. At the same
time, a new state $X(3872)$ was found by Belle \cite{BelleX} and CDF
\cite{CDF}.

This discovery has attracted much attention. As the state is just at
the $D \bar D^*$ threshold, it was immediately suggested
\cite{TornqvistX} that it might be a $D \bar D^*$ molecule bound by
pion exchange ("deuson"), considered long ago in \cite{TornqvistD}
and, much earlier, in \cite{VO} and \cite{RGG}. This requires
$1^{++}$ quantum numbers, and this assignment seems to be favoured
by the data \cite{abe, bauer,olsen}. The discovery channel is
$\pi^+\pi^- J/\psi$ with dipion most probably originating from the
$\rho$. Together with the observation of $X$ in the $\omega J/\psi$
channel \cite{Belleomega} this opens fascinating possibilities for
strong isospin violation, which is also along the lines of the
deuson model.

Other options for $X(3872)$ are under discussion in the literature, see
e.g. \cite{CP, BG, PS, Bugg}. The most obvious possibility of $X$ being a
$c
\bar c$ state seems to be ruled out by its mass: the state is too high to
be a $1D$ charmonium, and too low to be a $2P$ one \cite{BG}. This assumes
that we do know the spectrum of higher charmonia, namely, the
fine splittings and the role of coupling to $D$-meson pairs. It is the
latter issue which is addressed in the present paper.

The mechanism of open-flavour strong decay is not well-understood.
The simplest model for light-quark pair creation is the so-called
$^3P_0$ model, suggested many years ago \cite{Micu}. It assumes that
the pair is created with vacuum ($^3P_0$) quantum numbers uniformly
in space. The application of this model has a long history \cite{Orsay,
BO, KI}. Systematic studies \cite{Barnes1, Barnes2} of the decays of
light and strange quarkonia show that with a $^3P_0$ -type amplitude
calculated widths agree with data to within $25-40\%$. Recently the
charmonia decays \cite{Barnes3} and decays of $D$- and $D_s$-mesons
\cite{CS} were considered in the framework of the
$^3P_0$ model.

There exist also microscopic models of strong decays, which relate
the pair-creation interaction to the interaction responsible for the
formation of the spectrum, by constructing the current-current
interaction due to confining force and one-gluon exchange. Among
these is the Cornell model \cite{Eichten} which assumes that confinement
has Lorentz vector nature. The model \cite{ABarnes} assumes that the
confining interaction is the scalar one, while one-gluon-exchange
is, of course, Lorentz vector. Possible mechanisms of strong decays were
studied in the framework of Field Correlator Method (FCM) \cite{Simonov},
and an effective $^3P_0$ operator for open-flavour decay has emerged from
this study, with the strength computed in terms of FCM
parameters (string tension and gluonic correlation length).

Most of the above-mentioned papers are devoted to computing the widths,
and only a few consider the effects of virtual hadronic loops on the
spectra. The Cornell model \cite{Eichten} has presented a detailed
analysis of charmonia with coupling to $D$-mesons taken into account.
The recent update \cite{nEichten} of the Cornell model has presented
splittings caused by coupling to mesonic channels for $1D$ and $2P$ $c
\bar c$ levels, confirming the previous result: $X(3872)$ is well above
the range of $1D$ levels and well below the range of $2P$ ones.
The paper \cite{Barnes0} has reported first results for hadronic shifts
of lower charmonia due to mixing with $D$-meson pairs, calculated within
the $^3P_0$ model. The shifts appear to be alarmingly large.

Meanwhile, phenomenological coupled-channel models like \cite{port,
Tornqvist, Markushin, Pennington} accumulate experience on the
possibilities to generate nontrivial effects due to the coupling to
hadronic channels. As a recent example one should mention the
analyses \cite{Ds} of new $D_{sJ}$ states with masses considerably
lower than quark model predictions, and coupling to mesonic channels
being responsible for these anomalously low masses. It is
interesting to note that the coupled-channel calculations performed
in the framework of chiral Lagrangian approach \cite{ST} has arrived
at the same conclusions.

In this paper the coupled-channel model for charmonia levels is presented,
based on the nonrelativistic quark model for $c \bar c$ spectrum and
$^3P_0$-type model for pair-creation. In Section II the dynamics of
coupled channels is briefly outlined. Section III introduces the quark
model. Sections IV and V contain the results which are discussed in
Section VI. The paper ends with a short summary.

\section{Dynamics of coupled channels}

The details of coupled-channel model can be found e.g. in \cite{Eichten}, 
\cite{Ker}. Here I review the essentials.

In what follows the simplest version of coupled-channel model is
employed.
Namely, it is assumed that the hadronic state is represented
as
\be
|\Psi \rangle = \left(\sum_{\alpha}c_{\alpha}|\psi _{\alpha}\rangle\atop
\sum_i
\chi_i |M_1(i)M_2(i) \rangle \right),
\label{state}
\ee
where the index $\alpha$ labels bare confined states
$|\psi_{\alpha}\rangle$ with the probability amplitude $c_{\alpha}$, and
$\chi_i$ is the wave function in
the $i$-th two-meson channel $|M_1(i)M_2(i)\rangle$.
The wave function $|\Psi\rangle$ obeys the equation
\be
\hat {\mathcal H} |\Psi\rangle = M |\Psi\rangle,\quad
\hat {\mathcal H} = \left(\begin{array}{cc}
\hat{H}_c&\hat{V}\\
\hat{V}&\hat{H}_{M_1M_2}
\end{array}
\right),
\label{ham}
\ee
where $\hat {H}_c$ defines the discrete spectrum of bare states, with
$\hat {H}_c |\psi_{\alpha}\rangle = M_{\alpha} |\psi_{\alpha}\rangle$.
The part $\hat{H}_{M_1M_2}$ includes only the free-meson Hamiltonian, so
that the direct
meson-meson interaction (e.g., due to $t$- or $u$-channel exchange forces)
is neglected. The term $\hat {V}$ is responsible for dressing of the bare
states.

Consider one bare state $|\psi_0\rangle$ (the generalization to
multi-level case is straightforward).
The interaction part is given by the transition form factor
$f_i(\vep)$,
\be
\langle \psi_0|\hat {V}|M_{i1}M_{i2}\rangle = f_i(\vep_{i}),
\label{ff}
\ee
where $\vep_{i}$ is the relative momentum in $i$-th mesonic channel.
Then (\ref{ham}) leads to the  system of coupled equations for
$c_0(M)$ and $\chi_{i,M}(\vep_i)$:
\be
\left\{
\begin{array}{c}
\ds c_0(M)M_0 + \sum_i \int f_i(\vep)\chi_{i,M}(\vep)d^3p = Mc_0(M),\\[2mm]
\ds\left(m_{i1}+m_{i2}+\frac{p^2}{2\mu_{i}}\right)\chi_{i,M}(\vep_i) + c_0(M)f_i(\vep_i)
= M
\chi_{i,M}(\vep_i).
\end{array}
\right.
\label{system}
\ee
Here $\mu_i=\frac{m_{i1}m_{i2}}{m_{i1}+m_{i2}}$ is the reduced mass in the
system of
mesons with the masses $m_{i1}$ and $m_{i2}$, and $M_0$ is the
mass of the bare state. In what follows the formalism will be applied to
the system of charmed mesons, so the nonrelativistic kinematics is
employed in (\ref{system}). The fully relativistic version of
coupled-channel model is presented in \cite{Tornqvist}.

With the help of (\ref{system}) one easily calculates the $t$-matrix in
the mesonic system:
$$
t_{ik}(\vep_{i},\vep'_{k},M)=\frac{f_{i}(\vep_i)f_k(\vep'_k)}{M-M_0+g(M)},\quad
g(M)=\sum_i g_i(M),
$$
\be
g_i(M)=\int \frac{f_i(\vep)f_i(\vep)}{\frac{p^2}{2\mu_i}-E_i-i0} d^3p,
\label{t}
\ee
where $E_i=M-m_{i1}-m_{i2}$.
The quantity $g(M)$ is often called the hadronic shift of the bare state
$|\psi_0\rangle$, as the masses of physical states are defined, in
accordance with eq.(\ref{t}), from the equation
\be
M-M_0+g(M)=0.
\label{poles}
\ee

Let the eq.(\ref{poles}) have the solution
$M_B$ with $M_B$ smaller than the lowest mesonic threshold, so
there is
a bound state with the wave function
\be
|\Psi_B\rangle=\left(\cos\theta|\psi_0\rangle\atop
\sin\theta\sum_i\chi_{iB}(\vep_i)\rangle
 \right),\quad\langle\Psi_B|\Psi_B\rangle=1,\quad\cos\theta=\langle\psi_0|\Psi_B\rangle.
\label{wf}
\ee
Here $\sum_i\chi_{iB}(\vep_i)$ is normalized to unity, and $\cos\theta$
defines the
admixture of the bare state $|\psi_0\rangle$ in the physical state
$|\Psi_B\rangle$. The explicit expression for this admixture reads
\be
Z \equiv \cos^2\theta = \left(1+\sum_i\int
\frac{f_i(\vep)f_i^*(\vep')d^3p}{(\frac{p^2}{2\mu_{i}}+\epsilon_i)^2}
\right)^{-1}=
\left(\left.1+\frac{\partial g(M)}{\partial M}\right|_{M=M_B}\right)^{-1},
\label{Z}
\ee
$\epsilon_i=m_{i1}+m_{i2}-M_B, \epsilon_i>0$.
As far as I know, this $Z$-factor was first introduced by S.Weinberg in
\cite{SWein} many years ago as the field renormalization factor which
defines the probability to find the physical deuteron $|d \rangle$
in a bare elementary-particle state $|d_0 \rangle$, $Z= |\langle d_0|d
\rangle|^2$.

Even more detailed information is contained in the continuum counterpart
of the factor $Z$, the spectral density $w(M)$ of the bare state, given by
\be
w(M)=\sum_i w_i(M),\quad w_i(M)= 4\pi\mu_i p_i |c(M)|^2
\Theta(M-m_{i1}-m_{i2}),
\label{w}
\ee
where $c(M)$ is the probability amplitude to find the
bare
state in the continuum wave function $|\Psi \rangle_M$. With $c(M)$ found
from the system of equations (\ref{system}), one can calculate $w(M)$:
\be
w(M)=\frac{1}{2\pi i} \left (
\frac{1}{M-M_0+g^*(M)}-\frac{1}{M-M_0+g(M)} \right).
\label{disc}
\ee
As shown in \cite{bhm}, the normalization condition for the distribution
$w(M)$ follows from the completeness
relation for the total wave function (\ref{state}) projected onto bare
state channel, and reads:
\be
\int_{m_{01}+m_{02}}^{\infty} w(M)dM=1-Z,
\label{intw}
\ee
if the system possesses a bound state, and
\be
\int_{m_{01}+m_{02}}^{\infty} w(M)dM=1,
\label{intw1}
\ee
if there is no bound state ($m_{01}+m_{02}$ is the lowest
threshold). In the case of bound state
present, all the information on the factor $Z$ is encoded, due to eq.
(\ref{intw}), in the $w(M)$ too. On the other hand, the analysis in terms
of $w(M)$ can be performed in the case of resonance as well, as
exemplified in \cite{evidence}.

In the latter case the $t$-matrix poles are situated in the complex
plane. While the positions of the poles are the fundamental
quantities, another quantities are useful for practical purposes.
Namely, one defines the visible resonance mass $M_R$ from the
equation
\be M_R-M_0+{\rm Re}~g(M_R)=0,
\label{visiblemass}
\ee
and calculates the visible width as
\be
\Gamma=2\Re~{\rm
Im}~g(M_R),\quad\Re =\left(\left.1+\frac{\partial ~{\rm Re}~ g(M)}{\partial
M}\right|_{M=M_R}\right)^{-1}.
\label{visiblewidth}
\ee
Clearly this
brings the $t$-matrix into Breit-Wigner form, i.e. in the
form in which experimental data are usually delivered. In what follows the
factor $\Re$ will be called the renormalization factor.

There are some limitations of course, as not the every peak has the
Breit-Wigner shape. In the case of overlapping resonances the formulae
(\ref{visiblemass}) and (\ref{visiblewidth}) do not work.
The special case of near-threshold
$S$-wave resonance is not described by Breit-Wigner or Flatt{\`e} formula,
and the scattering length parametrization is more appropriate
\cite{Flatte}.

The quantities $Z$ and $w(M)$ are the ones of immediate relevance.
Indeed, there is no hope that, say, the elastic $D \bar D$
scattering will be measured some time. Our knowledge on mesonic
resonances
comes from external reactions, like $e^+e^-$ annihilation, $\gamma
\gamma $ collisions, $B$-meson decays etc. etc. Assuming that such
reactions proceed
via intermediate $q \bar q$ states, one obtains that the
cross-section is proportional to $w(M)$:
\be
\sigma(\to {\rm mesons})\propto\Gamma_{0r}~w(M),
\label{crossection}
\ee
where $\Gamma_{0r}$ is the
width of the bare state corresponding to the external reaction. Such
formulae were used in \cite{Eichten} to describe
the $e^+e^-$ annihilation into charmed mesons. In the limit of
narrow resonance eq.(\ref{crossection}) is reduced to the standard
Breit-Wigner formula
\be
\sigma(\to {\rm mesons}) \propto
\frac{1}{2\pi}\frac{\Gamma^0_r~\Gamma}{(M-M_0)^2+\frac{1}{4}\Gamma^2},
\label{BW}
\ee
where $\Gamma=2{\rm Im}\,g(M_0)$ is the (small) width
of the resonance. Similarly, for the bound state case the width
$\Gamma_r$ for a given reaction is renormalized as
\be
\Gamma_r=Z~\Gamma^0_r.
\label{Zleptwidth}
\ee

\section{The quark model}

This section specifies the form factors $f_i(\vep)$. The pair-creation
model
employed is the $^3P_0$ one, that is the pair-creation Hamiltonian is
the nonrelativistic reduction of
\be
H_q=g_q\int d^3x {\bar \psi_q} \psi_q,
\label{hcr}
\ee
for a given flavour $q$, but two important points make it
different from the model used in \cite{Barnes1,Barnes2,Barnes3}.

The
approach \cite{Barnes1,Barnes2,Barnes3} assumes that the pair creation
is flavour-independent, which yields for the constant $g_q$ the form
\be
g_q=\gamma~2m_q,
\label{gq}
\ee
where $\gamma$ is the effective strength of pair-creation. The factor
$2m_q$ implies enhancement of strange
quarks creation comparing to light quarks one. There are no fundamental
reasons to have such enhancement. Moreover, such factor is
absent in microscopical models of pair creation, like \cite{Eichten}
and \cite{ABarnes}. So, throughout the present study, I use the effective
strength $\gamma$ for the creation of light ($u$- and $d$-)
flavours, while for strange quarks the effective strength
$\gamma_s=\frac{m_q}{m_s}\gamma$ is used, where $m_q$ and $m_s$ are the
constituent masses of light and strange quarks correspondingly.

The authors of \cite{Barnes1,Barnes2,Barnes3} argue
that the assumption of flavour-independence gives a reasonably accurate
description of known decays. One should have in mind, however, that
the calculations \cite{Barnes1,Barnes2,Barnes3} are performed with the
so-called SHO wavefunctions, i.e. with the wave functions of harmonic
oscillator, and with the same oscillator parameter $\beta$ for all states.
This assumption looks implausible, as the behaviour of form factors
is defined by scales of wavefunctions, which, in turn, are defined by
quark model. 

In what follows the standard nonrelativistic potential model
is introduced, with the Hamiltonian
\be
H_0=\frac{p^2}{m_c}+V(r)+C,\quad V(r)=\sigma r - \frac{4}{3}
\frac{\alpha_s}{r},
\label{potential}
\ee
$m_c$ is the mass of charmed quark.
This Hamiltonian should be supplied by Fermi-Breit-type relativistic
corrections, including spin-spin, spin-orbit and tensor force, which cause
splittings in the $^{2S+1}L_J$ multiplets. In the first approximation
these splitting should be calculated as perturbations, using the
eigenfunctions
of the Hamiltonian (\ref{potential}). The same interaction $V(r)$ should
be used in spectra and wavefunction calculations of $D$ ($D_s$) mesons.

In the first approximation the
pair-creation amplitude is to be calculated with the eigenfunctions of the
zero-order Hamiltonian. Use of the SHO wavefunctions simplify these
calculations drastically. So the procedure adopted is to find the
SHO wavefunctions (of the form $\exp{(-\frac{1}{2}\beta^2r^2)}$ multiplied
by appropriate polynomials) for each orbital momentum $L$ and radial
quantum number $n$, with the effective value of oscillator parameter
$\beta$ for each $L$ and $n$, which reproduce the r.m.s. of the states.

I use the following set of
potential model parameters:
$$
\alpha_s=0.55,~\sigma=0.175~GeV^2,~m_c=1.7~GeV,C=-0.271~GeV,
$$
\be
~m_q=0.33~GeV,~m_s=0.5~GeV.
\label{parameters}
\ee

The spin-dependent force is taken in the form
\be
V_{sd}=V_{HF}+\frac{2\alpha_s}{m_c^2r^3}~\veL \cdot
\veS-\frac{\sigma}{2m_c^2r}~\veL \cdot \veS
+\frac{4\alpha_s}{m_c^2r^3}~T,
\label{SD}
\ee
where $\veL \cdot \veS$ and $T$ are spin-orbit
and tensor operators correspondingly, and $V_{HF}$ is the contact
hyperfine interaction,
\be
V_{HF}(r)=\frac{32\pi\alpha_s}{9m_c^2}~{\tilde \delta}(r)~\veS_{q} \cdot
\veS_{\bar q},
\label{HF}
\ee
where, following the lines of \cite{Barnes3}, Gaussian-smearing of the
hyperfine interaction is introduced,
\be
{\tilde \delta}(r)=\left(\frac{\kappa}{\sqrt{\pi}}\right)^3 e^{-\kappa^2 r^2},
\label{kappa}
\ee
with $\kappa=1.45$ GeV.

\begin{table}[t]
\caption{Masses and effective values $\beta$ (in units GeV).}
\begin{ruledtabular}
\begin{tabular}{cccc}
$nL$&$\beta(nL)$&State&Mass\\
\hline
$1S$&0.676&$1^3S_1$&3.264\\
&&$1^1S_0$&3.135\\
\hline
$2S$&0.485&$2^3S_1$&3.905\\
&&$2^1S_0$&3.850\\
\hline
$1P$&0.514&$1^3P_2$&3.773\\
&&$1^3P_1$&3.718\\
&&$1^3P_0$&3.631\\
&&$1^1P_1$&3.732\\
\hline
$2P$&0.435&$2^3P_2$&4.230\\
&&$2^3P_1$&4.181\\
&&$2^3P_0$&4.108\\
&&$2^1P_1$&4.192\\
\hline
$1D$&0.461&$1^3D_3$&4.051\\
&&$1^3D_2$&4.043\\
&&$1^3D_1$&4.026\\
&&$1^1D_2$&4.043\\
\end{tabular}
\end{ruledtabular}
\end{table}

The masses and effective values of oscillator
parameters $\beta$ for the model (\ref{parameters}) are listed in the
Table~I. The effective values of oscillator parameter for $D$-mesons are
$\beta_{D}=0.385$ GeV, and $\beta_{D_s}=0.448$ GeV.

One should not take the numbers given in last column too
seriously, especially for higher states, as the fine splittings are not
well-known, and the expression (\ref{SD}) is surely too naive.
Moreover, various much more sophisticated
approaches, which reproduce the splittings in $1P$ multiplet, give
different predictions for higher multiplets, as discussed in detail in
\cite{Alla}.

The $D$-meson masses taken are $M_D=1.867$
GeV, $M_{D^*}=2.008$ GeV, $M_{D_s}=1.969$ GeV,
$M_{D_S^*}=2.112$ GeV, so that the mass difference between neutral and
charged $D$-mesons is not taken into account. The pair-creation
strength for light quarks $\gamma=0.322$ is used. The $^3P_0$ amplitudes
are listed in the Appendix A.

\section{Lower charmonia and $D$-levels}

The single-level version of the coupled-channel
model is used in what follows, with the exception of $2^3S_1-1^3D_1$
levels.

All the physical charmonium masses below threshold are known. The hadronic
shifts and bare masses were calculated from the equation
\ref{poles},
\be
M_0=M_{phys}+\delta,\quad \delta=g(M_{phys})=\sum_ig_i(M_{phys}),
\label{shift}
\ee
where the sum is over mesonic channels $D \bar D$, $D \bar D^*$, $D^*
\bar D^*$, $D_s
\bar D_s$, $D_s \bar D_s^*$ and $D_s^* \bar D_s^*$ (an obvious shorthand
notation is
used here and in what follows: $D \bar D^* \equiv D \bar
D^* + \bar D D^*$, and $D_s \bar D_s^* \equiv D_s \bar
D_s^* + \bar D_s D_s^*$).

Besides, as the position of the
$1^{--}$ $\psi(3770)$ state is well-established, the bare mass of
$1^3D_1$ state was reconstructed by means of eq.\ref{visiblemass}, and the
visible width was calculated as (\ref{visiblewidth}).

The parameters of the underlying quark model (\ref{parameters}) and
the pair-creation strength $\gamma=0.322$ were
chosen to reproduce, with reasonable accuracy, the model masses of $1S$,
$1P$ and $2S$ states, and the width of $\psi(3770)$.

The results for bound states
are given in Table~II together with corresponding values of
$Z$-factors. The shifts are much smaller than in \cite{Barnes3}, but still 
substantial. 

\begin{table}[t]
\caption{Hadronic shifts (in units MeV) of charmonium
states below $DD$
threshold due to individual channels, total shifts $\delta$, bare
masses and
$Z$-factors.
The results for the $2^3S_1$ state do not take into account mixing with
the $1^3D_1$ state}
\begin{ruledtabular}
\begin{tabular}{c|ccccccc|c|c}
$n^{2S+1}L_J$&$DD$&$DD^*$&$D^*D^*$&$D_sD_s$&$D_sD^*_s$&$D^*_sD^*_s$&
$\delta$&$M_0$&$Z$\\
\hline
$1^3S_1$&11&42&69&5&19&31&177&3274&0.899\\
$1^1S_0$&0&59&55&0&26&25&165&3145&0.913\\
\hline
$1^3P_2$&25&64&82&8&22&27&228&3784&0.804\\
$1^3P_1$&0&70&91&0&22&32&215&3726&0.820\\
$1^3P_0$&29&0&118&8&0&43&198&3613&0.841\\
$1^1P_1$&0&87&76&0&29&27&219&3744&0.817\\
\hline
$2^3S_1$&21&60&87&45&16&27&216&3902&0.743\\
$2^1S_0$&0&83&71&0&24&22&200&3838&0.802\\
\end{tabular}
\end{ruledtabular}
\end{table}

Let me now discuss the $1^3D_1$ level, lying above $D \bar D$ threshold.
The bare mass is calculated to be $4.018$ GeV. The calculated width of the
$\psi(3770)$ is $25.5$ MeV, which compares well with the PDG value of
$23.6 \pm 2.7$ MeV \cite{PDG}. Note that it is the visible width,
while naive calculations would give
\be \Gamma_0=2~{\rm
Im}~g(M_R)=34.3 MeV.
\label{naivewidth}
\ee
So the effect of coupling
to mesonic channels on the width of $\psi(3770)$ is not small,
$\Re=0.743$.

The mass of $\psi(3770)$ is less than $100$ MeV higher than the mass of
$\psi'(3686)$, with $D \bar D$ threshold opening in between, so one could
in principle expect that these states are mixed due to coupling to mesonic
channels. This mixing is not large in the given model, as the mixing
disappears if the mass difference between various $D$-mesons is neglected
(see the corresponding spin-orbit recoupling coefficients listed in
Appendix A). The
relevant formulae for two-level mixing scheme are given in the Appendix
B, and here I quote the results. With physical masses of $\psi'(3686)$ and
$\psi(3770)$ the bare masses are reconstructed as $M(2^3S_1)=3.899$ GeV,
and $M(1^3D_1)=4.016$ GeV, so the masses are shifted only by a few MeV
due to the mixing.
The width of the $\psi(3770)$ becomes only $18.4$ MeV, but it is the
visible
width, and the deviation from the value $25.5$ MeV obtained without mixing
is mainly due to the illegitimate attempt to fit the system of two
overlapping states with a single-Breit-Wigner lineshape. The states indeed
do overlap, as shown at Fig.1, where the spectral densities of bare
$2^3S_1$ and $1^3D_1$ bare states are plotted. The lineshape of $1^3D_1$
is distorted due to the mixing, and the lineshape
of the $2^3S_1$ is drastically changed,
displaying, instead of small smooth background, a peak at the mass of
about $3760$ MeV. The $10$ MeV difference between the peak position and
the mass of $\psi(3770)$ is again due to the prescription of visible
width: single-Breit-Wigner approximation is not appropriate both
for $^3S_1$ and $^3D_1$ lineshapes.
As to the lower state $\psi'(3686)$, the admixture $Z_1$ of
the $2^3S_1$ bare state is $0.742$, while the admixture $Z_2$ of the bare
$1^3D_1$ state is $0.00063$.

\begin{figure}[ht]
\begin{tabular}{cc}
\epsfig{file=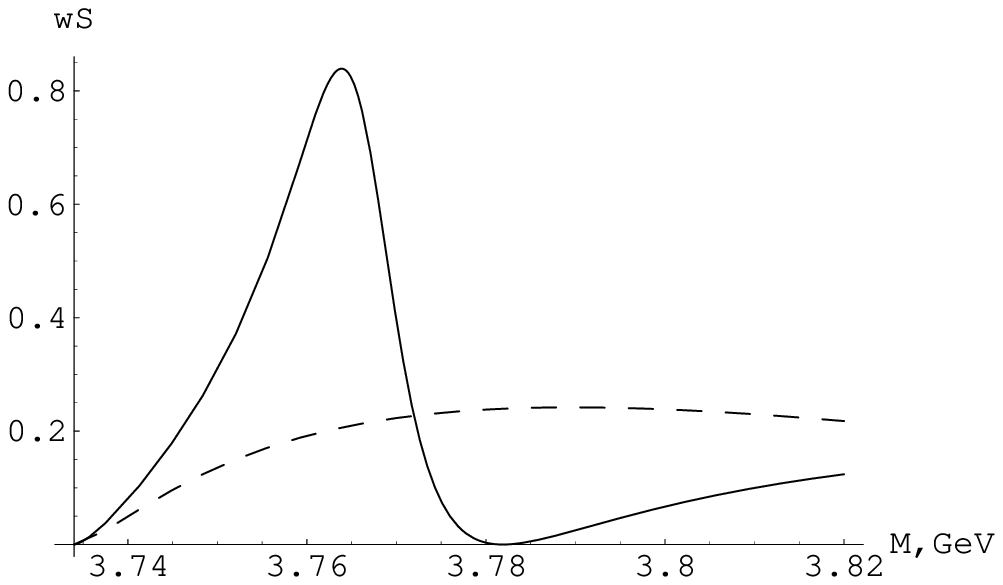,width=8cm}&\epsfig{file=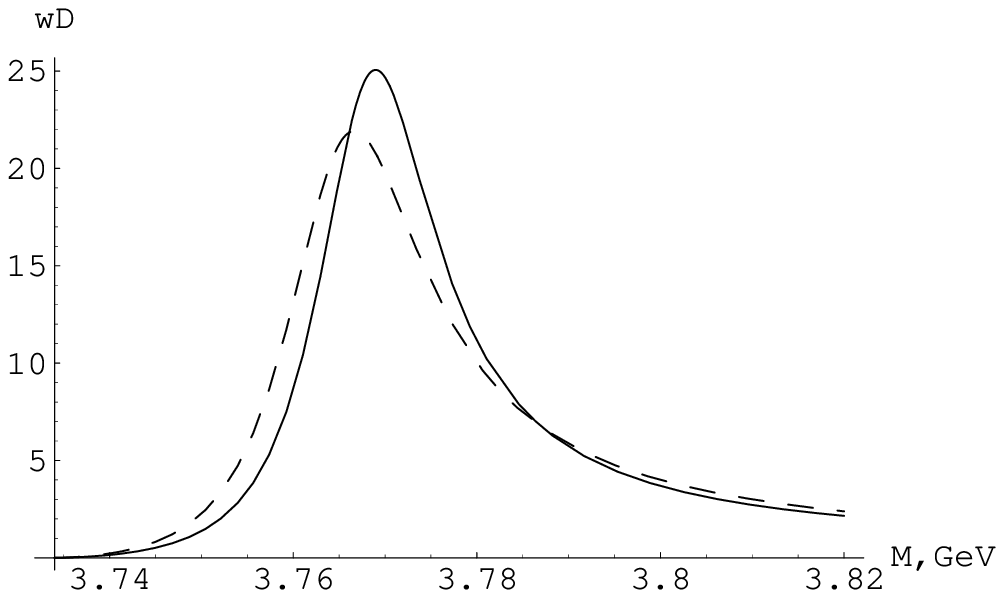,width=8cm}\\
a)&b)
\end{tabular}
\caption{a) Spectral density of $2^3S_1$ bare state with (solid line) and
without (dashed line) mixing. b) The same for the bare $1^3D_1$ state.}
\end{figure}

The calculated mass of bare $1^3D_1$ state is $4.016-4.018$ GeV. So the
value of $4.026$ GeV given in the Table~I looks quite acceptable.
No other $1D$-states are known, so in what follows I take, with
reservations mentioned in the previous section, the values of
bare masses from the Table~I.
With the bare $1^1D_2$ state having mass of $4.043$ GeV, the physical
state is a bound state,
\be
M(1^1D_2)=3.800~GeV,\quad Z(1^1D_2)=0.712.
\label{1D2}
\ee
Similarly,
\be
M(1^3D_2)=3.806~GeV,\quad Z(1^3D_2)=0.689
\label{3D2}
\ee
for the mass $4.043$ GeV of the bare $1^3D_2$ state.
The $1^3D_3$ state is allowed
to decay into $D \bar D$, but the width is extremely small, as the $D \bar
D$ system is in the $F$-wave:
\be
M(1^3D_3)=3.812~GeV,\quad \Gamma(1^3D_3) \approx 0.7~MeV,\quad \Re(1^3D_3)=0.717.
\label{3D3}
\ee

\section{$2P$-levels}

Coupled-channel effects do not cause dramatic changes for the charmonia
states discussed in the previous section. The situation with $2P$-levels
promises more, as $2P$-charmonia are expected to populate the mass
range of $3.90-4.00$ GeV, where more charmed meson channels start to
open, and some of these channels are the $S$-wave
ones.

The importance of $S$-wave channels follows from the eq.(\ref{t}). The form
factor $f(p)$ of the $S$-wave mesonic channel behaves as some constant at
small $p$, so the derivative of the hadronic shift $g(M)$ with respect to
the mass is large for the masses close to the $S$-wave threshold. As the
result,
hadronic shift due to the coupling to $S$-wave channel displays rather
vivid cusp-like near-threshold behaviour.

\begin{table}[t]
\caption{The masses and widths (in units of MeV), and renormalization
factors $\Re$ of $2P$ charmonium
states. The results for the $2^3P_0$ state are given for two different
values of bare mass as described in the text}
\begin{ruledtabular}
\begin{tabular}{cccccc}
State&Bare mass&Mass&Mode&Width&$\Re$\\
\hline
$2^1P_1$&4200&3980&$D \bar D^*$&50&0.615\\
\hline
$2^3P_2$&4230&3990&$D \bar D$&22&0.603\\
&&&$D_s \bar D_s$&1&\\
&&&$D \bar D^*$&45&\\
&&&total&68&\\
\hline
$2^3P_1$&4180&3990&$D \bar D^*$&27&0.543\\
\hline
$2^3P_0$&4108&3918&$D \bar D$&7&0.622\\
$2^3P_0$&4140&3937&$D \bar D$&8&0.393\\
\end{tabular}
\end{ruledtabular}
\end{table}

The physical masses and widths of $2P$ states were calculated with bare
masses given by Table~I, and the results
are listed in Table~III. Two different values
are given for the mass and width of the $2^3P_0$ state, for two
different choices of the bare mass, see below.
Looking at the numbers one would say that nothing
dramatic has happened due to $S$-wave thresholds. Indeed, all the shifts
are about $200$ MeV, and the renormalization factors are
about $0.5-0.6$. It is the behaviour of spectral density which
reveals the role of $S$-wave thresholds.

For the $2^1P_1$ case the $S$-wave
thresholds are $D \bar D^*$ with the spin-orbit recoupling coefficient
$C=-1/\sqrt{2}$ and multiplicity 4, the $D_s \bar D_s^*$ with
$C=-1/\sqrt{2}$
and multiplicity 2, the $D^* \bar D^*$ with $C=1$ and multiplicity 2, and
$D_s^* \bar D_s^*$ with $C=1$ and multiplicity 1. If the resonance is in
the mass range $3.90-4.00$ GeV (and it appears to be so), then the
relevant thresholds are $D \bar D^*$ and $D^* \bar D^*$.
The spectral density of the bare $2^1P_1$ state is shown at Fig.2. It
displays the relatively steep rise near $D \bar D^*$ threshold, and a
beautiful well-pronounced cusp due to the opening of  $D^* \bar D^*$
channel.

\begin{figure}
\centerline{\epsfig{file=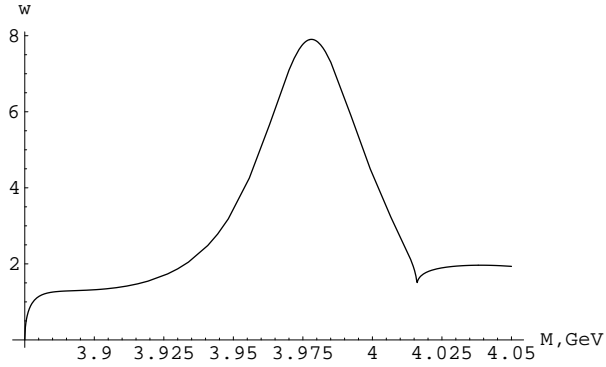, width=8cm}}
\caption{Spectral density of the bare $2^1P_1$ state.}
\end{figure}

In the $2^1P_1$ case the
$S$-wave strength is shared equally between $D \bar D^*$
and $D^* \bar D^*$ channels, while for the $2^3P_2$
case all $S$-wave strength is concentrated in the  $D^* \bar D^*$,
channel. As the result, the cusp due to the opening of  $D^* \bar D^*$
channel is more spectacular in the $2^3P_2$ case, as shown at Fig.3.

\begin{figure}
\centerline{\epsfig{file=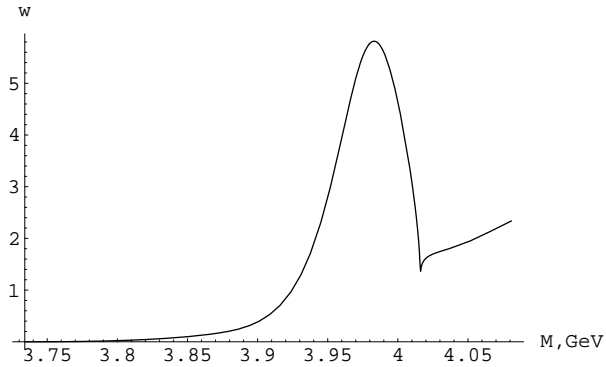, width=8cm}}
\caption{Spectral density of the bare $2^3P_2$ state.}
\end{figure}

The case of $2^3P_1$ is even more interesting. Here, similarly to the
$2^3P_2$ case, all the $S$-wave strength is concentrated in two channels,
$D \bar D^*$, $D_s \bar D_s^*$, and the multiplicity of the former is 4.
So the strongest $S$-wave threshold is the $D \bar D^*$ one, well below
the resonance. The behaviour of the $2^3P_1$ bare state spectral
density is shown at Fig.4, and is very peculiar: together with a clean and
relatively narrow resonance, there is a near-threshold peak, rising at
the flat background. The  $D \bar D^*$ scattering length appears to be
negative and large,
\be
a_{D \bar D^*}=-8~fm,
\label{a}
\ee
signalling the presence of virtual state very close to the
$D \bar D^*$ threshold, with the energy $\epsilon=0.32$ MeV. So the
coupling to mesonic channels has generated not only the resonance, but,
in addition, a virtual state very close to physical region.

\begin{figure}
\centerline{\epsfig{file=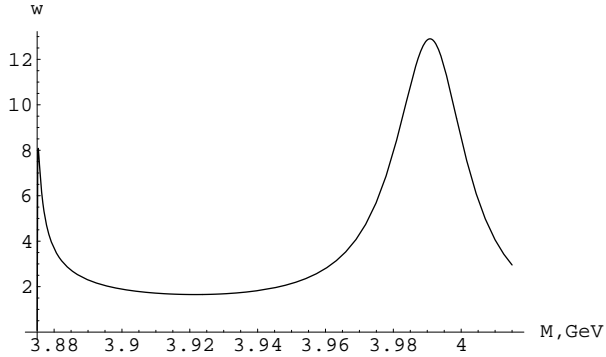, width=8cm}}
\caption{Spectral density of the bare $2^3P_1$ state.}
\end{figure}

\begin{figure}
\centerline{\epsfig{file=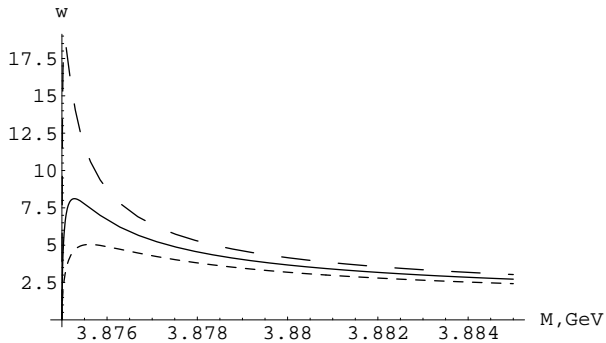, width=8cm}}
\caption{Near-threshold $2^3P_1$ spectral density for the mass of bare
state $M_0=4.180$ GeV (solid line), $M_0=4.190$ GeV (long-dashed line),
$M_0=4.170$ GeV (dashed line).}
\end{figure}

The near-threshold peak
should fade with the increase of bare state mass, and strengthen
otherwise. There are uncertainties in the
fine splitting estimates, so the mass of the bare $2^3P_1$ state could
easily be
about $10$ MeV larger or smaller. The dependence of spectral density
behaviour on the bare mass is shown at Fig.5. The peak becomes less
pronounced for the bare mass of $4.190$ GeV, but the scattering length
remains rather large, $a \approx -5.2$ fm, which corresponds to the
energy of
virtual state of about $0.76$ MeV (compare this with the scattering length
in the $1^{+-}$ channel, $|a| \approx 1$ fm). $10$ MeV decrease of the
bare state
mass leads to incredibly large scattering length, $a \approx -17.8$ fm,
and virtual state with the energy $0.065$ MeV.
Further decrease of bare state mass leads
to moving the state to the physical sheet, i.e. to appearance of the bound
state. This happens at the bare mass of about $4.160$ GeV, which seems, in
the present model, to be beyond acceptable range for fine splitting.
Similarly, the bound state appears if the
pair-creation strength is increased by several per cent.

\begin{figure}
\centerline{\epsfig{file=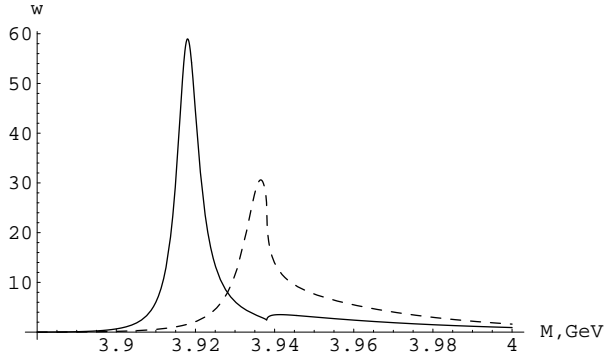, width=8cm}}
\caption{Spectral density of the bare $2^3P_0$ state for the
mass of bare
state $M_0=4.110$ GeV (solid line), $M_0=4.140$ GeV (dashed line).}
\end{figure}

The $2^3P_0$ level is a disaster, as it always happens with scalars. The
bare mass is considerably lower that the c.o.g., and the uncertainty in
the fine splitting estimate is large.
The $S$-wave $D^* \bar D^*$ and $D_s^* \bar D_s^*$
channels are too high. The $S$-wave $D \bar D$ channel is too low.
The only relevant
$S$-wave channel is $D_s \bar D_s$ ($C=\sqrt{\frac{3}{2}}$, multiplicity
1), and corresponding threshold is at
$3.938$ GeV, i.e. around the region where the resonance is expected. So,
depending on the position of the bare state, variety of spectral density
behaviour can be achieved, as shown at Fig.6, where spectral density is
plotted for $M_0(2^3P_0)=4.108$ GeV and $M_0(2^3P_0)=4.140$ GeV. Note that
two curves of Fig.6 correspond to two completely different
situations. The curve for $M_0(2^3P_0)=4.108$ GeV displays the
normal resonance behaviour with a tiny cusp due to the opening of the $D_s
\bar D_s$ channel. The curve for $M_0(2^3P_0)=4.140$ GeV is not
resonance-like at all. The formal exercise of calculating the
$\Re$-factor and visible width does not make much sense, and, as suggested
in \cite{Flatte}, such excitation
curve should be analysed in terms of scattering length approximation, and
not in terms of Breit-Wigner or Flatt{\`e} distributions.

\section{Discussion}

The quark model (\ref{parameters}) is not the result of the sophisticated
fit, it is rather a representative example.
A serious fit should include
proper treatment of fine and hyperfine splittings, as well as the mixing
of bare $2^3S_1$ and $1^3D_1$ states.
The calculations should be
performed with more realistic wavefunctions, and not with the SHO ones.
Relativistic corrections should be
taken into account in calculations of bare spectra, and more
realistic model should be used for $D$-meson wavefunctions. More
QCD-motivated model should be employed for the pair-creation Hamiltonian,
and loop integrals $g(M)$ are to be treated relativistically.
Nevertheless, there are gross features which are model-independent.

In various pair-creation models, the shifts within
the each orbital multiplet are approximately the same, and differ only due
to the mass difference of bare states in the multiplet and different
masses of charmed mesons. This is model-independent,
as the charmed quark is heavy, and the wavefunctions
within each $nL$ multiplet do not differ much. The same is true
for the wavefunctions of $D$ and $D^*$ (and $D_s$ and $D_s^*$) mesons due
to heavy quark spin symmetry.

In addition, the shifts for all states are more or less the
same. There are no {\it a priori} reasons for this, but the numbers given
in Table~II are stable up to overall multiplier $\gamma$ as soon as the
scales $\beta$ behave as expected from quark model, and the
value of $\gamma$ is constrained by experimental value of the $\psi(3770)$
width.

Indeed, the present results appear to be very similar to the ones given in
\cite{Eichten, nEichten}.
One might suggest that, from phenomenological point of view, the
effect of coupling to mesonic channels can be approximated by adding a
negative constant to the potential. But it is not the whole story, as
the coupling to charmed mesons generates the admixture of $D$-mesons in
the wavefunctions of bound states, which affects such quantities as
$e^+e^-$ widths (in accordance with eq\ref{Zleptwidth}) and the rates of
radiative transitions, as discussed in detail in \cite{Eichten}.

The $e^+e^-$ widths of $J/\psi$ and $\psi'(3686)$ are more or less
accurately described by Van Royen-Weisskopf formula with QCD correction,
so the renormalization (\ref{Zleptwidth}) is not harmless, even if it is
as mild as 10 $\%$ reduction required for $J/\psi$ by the results of
Table~II. For $\psi'(3686)$ $Z$-factor is 0.743, so renormalization is
larger.

In the nonrelativistic quark model the $e^+e^-$ width of the $^3D_1$ state
is zero. The mixing communicates the $e^+e^-$ width of the
$2^3S_1$ state to the $\psi(3770)$ region, with the $e^+e^- \to
D \bar D$ crossection given by eq.\ref{crossection} where the
spectral density of $2^3S_1$ state $w_S(M)$ is substituted.
It is reasonable to estimate the
$e^+e^-$ width of the $\psi(3770)$ using the peak value of $w_S(M)$,
which yields less than $1/3$ of the measured
value $0.26 \pm 0.04$ keV \cite{PDG}. Similar result was obtained in
\cite{Eichten}.
The  bare $1^3D_1$ state has the small $e^+e^-$ width of its own as
relativistic correction, and the bare $2^3S_1$ and $1^3D_1$ states are to
be mixed by tensor force.  The scale of the admixture required to
reproduce the relatively large $e^+e^-$ width of the $\psi(3770)$ is not
small, as shown in \cite{Rosner}. The coupling to charmed mesons is able to
explain only about $1/3$ of the observed $e^+e^-$ width of $\psi(3770)$,
so direct mixing between bare $2^3S_1$ and $1^3D_1$ bare levels is
still needed, which would reduce the
$e^+e^-$ width of the $\psi'$ further. While the problem of leptonic
widths is an open problem for coupled-channel model, it is clear that
the values of $Z$-factors considerably smaller than given in Table~II
would
destroy fragile agreement with the data on $e^+e^-$ widths achieved by
quark model practitioners.

The $2^3S_1$--$1^3D_1$ mixing due to coupling to meson channels is small,
and this is also model-independent, as it vanishes when the masses and
wavefunctions of $D$-mesons are taken to be the same. Note, however,
that both $\psi(3770)$ and $\psi'$ states contain considerable admixture
of four-quark component of their own, in the form of various $D$-meson
pairs. Thus, if the mass difference between charged and neutral $D$-mesons
is taken into account, some isospin violation could be generated. It is
argued in \cite{Voloshincc} that a small admixture of $I=1$ component is
needed to explain some discrepancies in the observed properties of
$\psi(3770)$ and $\psi'$. The question of whether such admixture can be
generated by coupled-channel effects certainly deserves attention.

The shifts of $D$-levels are more or less the same, the
relevant thresholds are $P$-wave ones, and the $2^3S_1$--$1^3D_1$ mixing
does not affect the shift of the bare $^3D_1$ level drastically. So the
combined effect of fine splitting and splitting due to coupled-channel
effects on other $D$-levels is not large. In particular, it means that,
with the mass of $\psi(3770)$ as experimental input, the physical $^3D_2$
or $^3D_3$
state cannot be placed as high as $3.872$ GeV and, therefore, cannot be
identified as $X(3872)$, unless something drastic happens with fine
splittings in the $1D$-multiplet.

Due to the presence of $S$-wave thresholds, the situation with $2P$-levels
is more interesting. The coupling to $D^* \bar D^*$ channels generates
pronounced
cusps in the spectral densities of bare $2^1P_1$ and $2^3P_2$ levels, and
the coupling to $D \bar D^*$ channel
generates the strong threshold effect in the $1^{++}$ wave. Within the
given model, it is a
virtual state with the energy less than $1$ MeV. The mechanism of
generating such a state is quite peculiar: it is one and a same bare
$2^3P_1$  state, which gives rise both to the $1^{++}$ resonance with the
mass of about $3990$ Mev, and a virtual state at the $D \bar D^*$
threshold, i.e. where the $X(3872)$ is observed. Due to the presence of
strong $S$-wave threshold, the hadronic
shift appears to be large enough to destroy the one-to-one correspondence
between bare and physical states, as widely discussed in
connection with light scalar mesons \cite{Tornqvist,Pennington}.

To what extent this prediction is robust? Changes of the
underlying quark model parameters or of the value of pair-creation
strength can shift this extra state either to the physical sheet or away
from the physical region. In the latter case, however, the $D
\bar D^*$ scattering length remains large.
One should have in mind that if such
dynamical generation of extra state at $D \bar D^*$ threshold is possible
in the charmonia, the
$1^{++}$ channel is the most appropriate place for this phenomenon. The
latter statement is model independent.

First, note that the scalar charmonium does not decay into $D \bar
D^*$ at all, and the tensor one decays into $D \bar D^*$ in the
$D$-wave. As to $1^{++}$ and $1^{+-}$ levels, they both have the
desired $S$-wave decay mode. Apply now the heavy quark spin
selection rule \cite{Voloshinspin}, which suggests that the spin of
a heavy quark pair is conserved in the decay. The $S$-wave decay
mode comes from the four-quark state $c \bar c q \bar q$ with all
relative angular momenta equal to zero. Then the total angular
momentum $J=1$ can result only from the quark spins. The combination
of $S_{c \bar c}=1$ and $S_{q \bar q}=1$ is $C$-even, while the
combination of $S_{c \bar c}=0$ and $S_{q \bar q}=1$ is $C$-odd. It
is a simple algebra exercise to show that, symbolically, 
\be 
(c \bar c)_{S=1} \otimes (q \bar q)_{S=1} \rightarrow \frac{1}{\sqrt{2}} 
(D \bar D^* + \bar D D^*), 
\label{even} 
\ee 
and 
\be 
(c \bar c)_{S=0} \otimes (q \bar q)_{S=1} \rightarrow -\frac{1}{2} (D^* 
\bar D - D
\bar D^*)+ \frac{1}{\sqrt{2}}D^* \bar D^*, 
\label{odd} 
\ee 
and,
independently of the pair-creation model, all the $S$-wave strength
of the $^3P_1$ decay is concentrated in the $D \bar D^*$ channel,
while in the $^1P_1$ decay it is shared equally between $D \bar D^*$
and $D^* \bar D^*$. Thus the threshold attraction in the $1^{++}$ $D
\bar D^*$ channel is always much stronger than in the $1^{+-}$ one.
It would be interesting to see if the pair-creation model of
\cite{Eichten} is able to generate large $D \bar D^*$ scattering
length.

 The model \cite{SwansonX}
contains the detailed analysis of the $X(3872)$ as a state bound
both by pion exchange and quark exchange in the form of transitions
$D \bar D^* \to J/\psi~\rho, J/\psi~\omega$, with the latter
contributions being important for the binding. In fact, this model
has predicted the $J/\psi~\omega$ decay mode of the $X(3872)$, and,
after observation \cite{Belleomega} of the decay $X(3872)
\rightarrow \pi^+\pi^-\pi^0 J/\psi $, it has become almost official
model of $X(3872)$. This is challenged by preliminary data from
Belle \cite{olsen} on large $D^0D^0\pi^0$ rate, more than ten times
larger than $\pi^+\pi^-J/\psi$ one, while the model \cite{SwansonX}
claims the opposite.

One could question validity of the naive quark-exchange model.
Besides, the pion exchange is definitely attractive in the $1^{++}$
channel, but there are uncertainties in the actual calculations; the
details of binding depend on the cutoff scale $\Lambda$, as
recognized in \cite{SwansonX}. The attraction found in the
coupled-channel model is large, and could help binding without
large quark-exchange kernels, and, correspondingly, without large
$\pi^+\pi^-J/\psi$ rate.

From practical point of view, the wavefunctions of both models are
not very distinguishable. Indeed, the near-threshold virtual state
of the coupled-channel model owes its existence to the bare $1^{++}$
state, but the near-threshold admixture of the bare state in the 
wavefunction is
extremely small, as seen from Fig.4 (recall that the spectral
density is normalized to unity). So the decays like $D \bar D \pi$
and $D \bar D \gamma$ would proceed via $D^*$ decays, as described
in \cite{Voloshindecay}. As to short-distance decays and exclusive
production, the rates of these are governed by large scattering
length. This phenomenon was called low-energy universality in
\cite{Braaten1}. Consider, for example, the near-threshold
production of $D \bar D^*$ pairs in the reaction $B \to D \bar
D^* K$. As explained in Section II, the $D \bar D^*$ invariant mass
distribution is proportional to $w(M)$ in the coupled channel model,
with the lineshape plotted at Fig.5. Now compare these curves with
the ones presented in \cite{Braaten2, Braaten3} with the scattering
length approximation for the $D \bar D^*$ amplitude, and observe
that the low-energy universality indeed takes place.

The lineshape for $D \bar D^*$ production depends only on the
modulus of scattering length, and the cases of bound state and
virtual state are not distinguishable. If some inelastic channel
like $\pi \pi J/\psi$ is present, then, as shown in \cite{Braaten3},
the lineshape for this channel depends on whether there is a bound
state or virtual state (cusp). The area under the cusp is much
smaller than the area under the resonance, so, in principle, the
data could distinguish between $X(3872)$ as a bound state or a
virtual state.

To conclude the discussion of coupled-channel virtual state as
$X(3872)$ I note that, with $7$ MeV difference between $D^0 \bar
D^{*0}$ and $D^+ \bar D^{*-}$ thresholds, the coupled-channel model
is able to generate considerable isospin violation. The coupled-channel 
calculations which resolve the mass difference between charged and neutral 
$D$-mesons are in progress now.

Last year yet another new state was reported \cite{BelleY} as an
enhancement in $\omega J/\psi$ mode, with the mass of $3941$ MeV and
the width of about $90$ MeV ($Y(3940$). As there are two $1^{++}$
physical states per one bare $2^3P_1$, it is tempting to identify
the $1^{++}$ resonance with the $Y(3940)$ state, following the
suggestion of \cite{Bugg}. Nevertheless, with the given set of quark
model parameters the resonance is $40-50$ MeV higher than $3940$
MeV, and is narrow.

One might suggest the $0^{++}$
assignment for $Y(3940)$, as the coupled-channel model places the scalar
just at the right place. However, as the $D_s \bar D_s$ channel is opening
at $3938$ MeV, the width of the state cannot be as large as observed $90$
MeV independently of what channel saturates it, unless, for some strange
reasons, the state couples weakly to $D_s \bar D_s$.
It is similar to $a_0(980)/f_0(980)$ case, where, in spite of
large coupling to light pseudoscalars ($\pi\pi$ or $\pi\eta$), the visible
width is small due to strong affinity to $K \bar K$ threshold.

\section{Conclusions}
The spectrum of charmonium states below $4$ GeV is calculated taking into
account coupling to the pairs of lowest $D$ and $D_s$ mesons. The analysis
is performed within the $^3P_0$ model for light-quark pair-creation. It
appears that quite moderate modification of quark model parameters is
needed to describe charmonia below $D \bar D$ threshold, while coupling
to open charm does not cause drastic effects on $1D$-levels. This is in
contrast to $2P$-levels, where opening of strong $S$-wave channels
leads to pronounced threshold effects.

In particular, coupling of the bare $2^3P_1$ state to $D \bar D^*$ channel
generates, together with the $1^{++}$ resonance with the mass of $3990$
MeV, a near-threshold virtual state with the energy of about $0.3$ MeV,
which corresponds to the extremely large  $D \bar D^*$
scattering length $a \approx - 8$ fm, with the possibility to identify
this state with $X(3872)$. The admixture of the bare $c \bar c$ state in 
the near-threshold wavefunction
is very small, so it is essentially the $D \bar
D^*$ state. The $1^{++}$ channel appears to be the only one where such
state can be formed.

\begin{acknowledgments}
I am grateful to Yu.A.Simonov for useful discussions.

This research is supported by the grant
NS-1774.2003.2, as well as of the
Federal Programme of the Russian Ministry of Industry, Science, and
Technology No 40.052.1.1.1112.
\end{acknowledgments}

\begin{center}
{\bf Appendix A}
\end{center}
\setcounter{equation}{00}
\renewcommand{\theequation}{A.\arabic{equation}}

The $^3P_0$ form factors are defined following the lines of
\cite{ABarnes,Barnes1}, adapted for charmonium transitions. Let the
$c \bar c$ meson $A$ decay to $q \bar c$ meson $B$ and $c \bar q$
meson $C$. Then the spin-space part of the amplitude in the c.m.
frame of the initial meson $A$ is given by \be I(\vep)=\int d^3k
\phi_A(\vek-\vep)<s_qs_{\bar q}|{\hat
O}(\vek)|0>\phi_B^*(\vek-r_q\vep) \phi_C^*(\vek-r_q\vep),
\label{amplitude} \ee $\phi_A$ is the wavefunction of the initial
meson in the momentum space, $\phi_B$ and $\phi_C$ are the
wavefunctions of the final mesons $B$ and $C$, $\vep=\vep_B$,
$r_q=\frac{m_q}{m_q+m_c}$, $m_c$ is the mass of charmed quark, $m_q$
is the mass of light quark. ${\hat O}(\vek)$ is the $^3P_0$
operator: \be {\hat O}(\vek)=-2\gamma ~\sigma \cdot \vek. \ee

The form factors for the transition $A \to B C$ are
given below in the $lS$ basis, where $l$ is the
orbital momentum in the final meson system, and $\vec S = \vec J_B +\vec
J_C $ is the total spin of the mesons $B$ and $C$. In the narrow width
approximation these form factors $f_{lS}$ define the partial widths
$\Gamma_{A \to B C}$ as
\be
\Gamma_{A \to BC}=2~{\rm Im}~g_{BC}(M_A)=2\pi p_{BC} \mu_{BC}
\sum_{lS}|f_{lS}|^2.
\label{Gamma}
\ee

The form factors calculated with SHO wavefunctions take the form
\be
f_{lS}=\frac{\gamma}{\pi^{1/4}\beta_A^{1/2}}\exp \left(
-\frac{p^2(r_q-1)^2}{\beta_B^2+2\beta_A^2} \right) ~{\cal P}_{lS},
\label{f}
\ee
where $\beta_A$ and $\beta_B=\beta_C$ are the oscillator parameters of
initial meson $A$ and final mesons $B$ and $C$, $\gamma$ is the
pair-creation strength, and $r_q=m_q/(m_c+m_q)$, $m_c$ is the mass of
charmed quark, $m_q$ is the mass of light quark. The polynomial ${\cal
P}_{LS}$ is a channel-dependent one:
\be
{\cal P}_{lS}=f_{l}~C_{lS},
\label{P}
\ee
where $C_{lS}$ are spin-orbit recoupling coefficients for specific mesonic
channels, and $f_l$ are:

\be
f_{P}(1S \to 1S +1S)=-\frac{2^{3}}{3^2}
\frac{\lambda\beta^3}{\beta_B^3\beta_A}p
\label{1S}
\ee

\be
f_P(2S \to 1S
+1S)=\frac{2^{5/2}}{3^{3/2}}
\frac{\beta^3}{\beta_A\beta_B^3}p
\left(-\lambda+\left(\frac{10}{9}\lambda-\frac{4}{9}\right)\frac{\beta^2}{\beta_A^2}+
\frac{2p^2}{3\beta_A^2}\lambda(\lambda-1)^2 \right)
\label{2S}
\ee

\be
f_S(1P \to 1S +1S)=
\frac{2^4}{3^{5/2}}\frac{\beta^5}{\beta_B^3\beta_A^2}
\Big (1+\lambda(\lambda-1)\frac{p^2}{\beta^2} \Big)
\label{1PS}
\ee

\be
f_D(1P \to 1S
+1S)=-\frac{2^4}{3^2
\cdot
5^{1/2}}\frac{\lambda(\lambda-1)\beta^3}{\beta_A^2\beta_B^3}p^2
\label{1PD}
\ee

\be
f_P(1D \to 1S
+1S)=
-\frac{2^{11/2}}{3^3}\frac{(\lambda-1)\beta^5}{\beta_A^3\beta_B^3}p
\Big(1+\frac{3}{5}\lambda(\lambda-1)\frac{p^2}{\beta^2} \Big)
\label{1DP}
\ee

\be
f_F(1D \to 1S +1S)=
-\frac{2^4}{3^{3/2}
\cdot 5^{1/2} \cdot 7^{1/2}}
\frac{\lambda(\lambda-1)^2\beta^3}{\beta_A^3\beta_B^3}
p^3
\label{1DF}
\ee

$$
f_S(2P \to 1S +1S)=
$$
\be
\frac{2^{7/2} \cdot
5^{1/2}}{3^{5/2}}\frac{\beta^5}{\beta_B^3\beta_A^2}
\Big
(1-\frac{2\beta^2}{3\beta_A^2}+\lambda(\lambda-1)\frac{p^2}{\beta^2}
-\frac{2p^2}{3\beta_A^2}
(\lambda-1)(2\lambda-1)-\frac{2p^4}{5\beta^2\beta_A^2}\lambda(\lambda-1)^3
\Big )
\label{2PS}
\ee

$$
f_D(2P \to 1S +1S)=
$$
\be
-\frac{2^{7/2}}{3^2}\frac{\beta^3}{\beta_A^2\beta_B^3}p^2 \Big (
\lambda(\lambda-1)-\frac{2\beta^2}{15\beta_A^2}(\lambda-1)(7\lambda-2)-
\frac{2p^2}{5\beta_A^2}\lambda(\lambda-1)^3 \Big )
\label{2PD}
\ee

Here
\be
\lambda=\frac{\beta_B^2+2r_q\beta_A^2}{\beta_B^2+2\beta_A^2},
\label{lambda}
\ee
and
\be
\beta^2=\frac{3\beta_A^2\beta_B^2}{\beta_B^2+2\beta_A^2}.
\label{beta}
\ee

For the transitions to strange mesons, one should replace $r_q$ by
$r_s=m_s/(m_c+m_s)$, and insert the multiplier $\frac{m_q}{m_s}$,
as explained in the main text.

The decays of light mesons were considered in \cite{Barnes1}, which 
corresponds to
$r_q=1/2$. In the case of $\beta_A=\beta_B=\beta$, and $r_q=1/2$ the
expressions for the amplitudes listed above are equal to
those of \cite{Barnes1} up to the factor of
$1/2$. In general, there are two graphs with different topologies
which contribute to the $^3P_0$ amplitude, and the sum of both is
quoted in \cite{Barnes1}, while in actual calculations each graph
contributes with the individual flavour factor. In the case of
charmonia transitions only one graph contributes, so that the
amplitude for the transition into given charge channel is equal to
(\ref{f}) with the flavour factor of unity.

The mass difference between neutral and charged mesons is not taken into
account, so the sum over charge states is equivalent to introducing the
multiplicity factor 2 for $D \bar D$ and $D^* \bar D^*$ channels, and 4
for $D \bar D^*$ channel. The multiplicity factor for $D_s \bar D_s$ and
$D_s^* \bar D_s^*$ is 1, and it is 2 for $D_s \bar D_s^*$ channel.

Spin-orbit recoupling coefficients are tabulated in the Appendix A of
\cite{Barnes1}. In the single-level coupled-channel calculations the
squares of spin-orbit recoupling coefficients are needed,
given in Table~IV. Note that the multiplicity factors for $D \bar D^*$
channels are twice as large as of $D \bar D$ and $D^* \bar D^*$ ones.

\begin{table}[t]
\caption{The sums $\sum_S |C_{lS}|^2$ for given initial and final states}
\begin{ruledtabular}
\begin{tabular}{c|cc|cc|cc}
State&$D \bar D$&&$D \bar D^*$&&$D^* \bar D^*$&\\
&$l=L-1$&$l=L+1$&$l=L-1$&$l=L+1$&$l=L-1$&$l=L+1$\\
\hline
$^3S_1$&-&1&-&2&-&7\\
$^1S_0$&-&0&-&3&-&6\\
\hline
$^1P_1$&0&0&1/2&5/3&1&10/3\\
$^3P_2$&0&1&0&3/2&2&8/3\\
$^3P_1$&0&0&1&5/6&0&5\\
$^3P_0$&3/2&0&0&0&1/2&20/3\\
\hline
$^1D_2$&0&0&1/4&7/5&1/2&14/5\\
$^3D_3$&0&1&0&4/3&1&29/15\\
$^3D_2$&0&0&3/8&14/15&1/4&56/15\\
$^3D_1$&5/12&0&5/24&0&1/6&28/5\\
\end{tabular}
\end{ruledtabular}
\end{table}

The case of $2^3S_1$--$1^3D_1$ mixing requires explicit expressions for
spin-orbit recoupling coefficients, as relative signs are important in the
two-level mixing scheme. These are given below:

$$
C_{10}(^3S_1 \to ^1S_0+^1S_0)=1
$$
$$
C_{11}(^3S_1 \to ^3S_1+^1S_0)=-\sqrt{2}
$$
$$
C_{10}(^3S_1 \to ^3S_1+^3S_1)=\sqrt{\frac{1}{3}}
$$
\be
C_{12}((^3S_1 \to ^3S_1+^3S_1)=-\sqrt{\frac{20}{3}}
\label{3s1}
\ee

\bigskip

$$
C_{10}(^3D_1 \to ^1S_0+^1S_0)=-\sqrt{\frac{5}{12}}
$$
$$
C_{11}(^3D_1 \to ^3S_1+^1S_0)=-\sqrt{\frac{5}{24}}
$$
$$
C_{10}(^3D_1 \to ^3S_1+^3S_1)=-\frac{\sqrt{5}}{6}
$$
$$
C_{12}(^3D_1 \to ^3S_1+^3S_1)=\frac{1}{6}
$$
\be
C_{32}(^3D_1 \to ^3S_1+^3S_1)=-\sqrt{\frac{28}{5}}
\label{3d1}
\ee

\begin{center}
{\bf Appendix B}
\end{center}
\setcounter{equation}{00}
\renewcommand{\theequation}{B.\arabic{equation}}

The formulae necessary to describe the $2^3S_1$--$1^3D_1$ mixing are
collected here.

Two sets of form factors, $f_S$ and $f_D$, are introduced, which describe
the transitions between mesons and $2^3S_1$, $1^3D_1$ levels. The
system of coupled channel equation similar to (\ref{system}) leads to the
$D \bar D$ $t$-matrix:
\be
t(\vep,\vep',M)=\sum_{\mu,\nu}f_{\mu,D \bar D}(\vep)
\tau_{\mu\nu}(M) f_{\nu,
D \bar D}(\vep'),
\label{tmix}
\ee
where sum is over $S$ and $D$ states, and
$$
\tau_{SS}(M)=\frac{M-M_D+g_{DD}(M)}{\Delta(M)},
$$
$$
\tau_{DD}(M)=\frac{M-M_S+g_{SS}(M)}{\Delta(M)},
$$
\be
\tau_{DS}(M)=\tau_{SD}(M)=-\frac{g_{SD}(M)}{\Delta(M)},
\label{tau}
\ee
\be
\Delta(M)=[M-M_S+g_{SS}(M)][M-M_D+g_{DD}(M)]-g_{SD}^2,
\label{delta}
\ee
$M_S$ and $M_D$ are the masses of bare states, and
\be
g_{\mu\nu}(M)=\sum_i g_{i,\mu\nu}(M),\quad g_{i,\mu\nu}(M)=\int
\frac{f_{\mu,i}(\vep)f_{\nu,i}(\vep)}{\frac{p^2}{2\mu_i}-E_i-i0} d^3p,
\label{g}
\ee
the index $i$ labels mesonic channels.

The visible physical masses are defined, similarly to (\ref{visiblemass}),
from the equation
\be
{\rm Re}~\Delta(M_{phys})=0.
\label{mixmass}
\ee
There are two solutions of this equation, $M_b$ below $D \bar D$
threshold,
corresponding to the bound state, and $M_a$ above threshold corresponding
to the resonance. The visible width is defined for the latter as
$$
\Gamma=2\Re~{\rm Im}~\Delta(M_a),
$$
$$
\Re^{-1}=
[1+d_{SS}(M_a)][M_a-M_D+{\rm Re}g_{DD}(M_a)]
$$
$$
+[1+d_{DD}(M_a)]
[M_a-M_S+{\rm Re}g_{SS}(M_a)]
-2{\rm Re}g_{SD}(M_A)d_{SD}(M_A),
$$
\be
d_{\mu\nu}(M)=\frac{\partial~{\rm Re}~g_{\mu\nu}(M)}{\partial M}.
\label{mixwidth}
\ee

The probabilities to find bare states in the wavefunction of bound state
are
\be
Z_S=[M_b-M_D+g_{DD}(M_b)]{\cal Z}^{-1},\quad Z_D=[M_b-M_S+g_{SS}(M_b)]{\cal
Z}^{-1},
\label{mixz}
\ee
and ${\cal Z}^{-1}$ is obtained from $\Re^{-1}$ given in (\ref{mixwidth})
with the replacement $M_a \to M_b$. The spectral densities of bare states
are:
\be
w_S(M)=\frac{1}{2\pi i}\Big(\frac{M-M_D+g^*_{DD}}{\Delta^*(M)}-
\frac{M-M_D+g_{DD}}{\Delta(M)} \Big ),
\label{ws}
\ee
and
\be
w_D(M)=\frac{1}{2\pi i}\Big(\frac{M-M_S+g^*_{SS}}{\Delta^*(M)}-
\frac{M-M_S+g_{SS}}{\Delta(M)} \Big ),
\label{wd}
\ee


\begin{thebibliography}{99}
\bibitem{Belle} S.-K. Choi {\it et al}. [Belle Collaboration], Phys. Rev.
Lett. 89, 102001 (2002) [arXiv:hep-ex/0206002].
\bibitem{Babar} B. Aubert {\it et al}. [BABAR Collaboration], Phys. Rev.
Lett. 92, 142002 (2004) [arXiv:hep-ex/0311038].
\bibitem{Cleo1} D. M. Asner {\it et al}. [CLEO Collaboration], Phys. Rev.
Lett. 92 142001 (2004) [arXiv:hep-ex/0312058].
\bibitem{FNAL} C. Patrignani [FNAL-E835 Collaboration], presented at BEACH
2004 [arXiv:hep-ex/0410085].
\bibitem{Cleo2} A. Tomaradze [CLEO Collaboration], presented at GHP2004
[arXiv:hep-ex/0410090].
\bibitem{BelleX} S. K. Choi {\it et al}. [Belle Collaboration], Phys. Rev.
Lett. 91, 262001 (2003) [arXiv:hep-ex/0309032].
\bibitem{CDF} D. Acosta  {\it et al}. [CDF II Collaboration], Phys. Rev.
Lett. 93, 072001 (2004)  [arXiv:hep-ex/0312021].
\bibitem{TornqvistX} N. A. Tornqvist, Phys. Lett. B590, 209 (2004)
[arXiv:hep-ph/0308277], [arXiv:hep-ph/0402237].
\bibitem{TornqvistD} N. A. Tornqvist, Phys. Rev. Lett. 67, 556 (1991)
\bibitem{VO} M. B.Voloshin and L. B. Okun, JETP Lett. 23, 333 (1976)
[Pisma Zh. Eksp. Teor. Fiz. 3, 369 (1976)]
\bibitem{RGG} A. de Rujula, H. Georgi, and S. L. Glashow, Phys. Rev. Lett.
38, 317 (1977)
\bibitem{abe} K. Abe, arXiv:hep-ex/0505038.
\bibitem{bauer} G. Bauer, arXiv:hep-ex/0505083.
\bibitem{olsen} S. L. Olsen, talk given at APC meeting 2005
[http://belle.kek.jp/belle/talks/aps05/olsen.pdf].
\bibitem{Belleomega} K. Abe {\it et al}. [Belle Collaboration],
[arXiv:hep-ex/0408116].
\bibitem{CP}F. E. Close and P. R. Page, Phys. Lett B578, 210 (2003)
[arXiv:hep-ph/0309253].
\bibitem{BG} T. Barnes ans S. Godfrey, Phys. Rev. D69 , 054008 (2004)
[arXiv:hep-ph/0311162].
\bibitem{PS} S. Pakvasa and M. Suzuki, Phys. Lett. B579, 67 (2004)
[arXiv:hep-ph/0309294].
\bibitem{Bugg} D. V. Bugg, Phys. Rev. D71, 016006 (2005)
[arXiv:hep-ph/0410168].
\bibitem{Micu}L. Micu, Nucl. Phys. B10, 521 (1969)
\bibitem{Orsay} A. Le Yaouanc, L. Oliver, O. Pene and J. Raynal, Phys.
Rev. D8, 223 (1973)
\bibitem{BO} G. Busetto and L. Oliver, Z. Phys. C20, 247 (1983)
\bibitem{KI} R. Kokoski and N. Isgur, Phys. Rev. D35, 907 (1987)
\bibitem{Barnes1}T. Barnes, F. E. Close, P. R. Page, E. S. Swanson,
Phys. Rev. D55 4157 (1997) [arXiv:hep-ph/9609339].
\bibitem{Barnes2} T. Barnes, N. Black, P.R. Page, Phys. Rev.
D68 054014 (2003) [arXiv:nucl-th/0208072].
\bibitem{Barnes3} T. Barnes, S. Godfrey and E. S. Swanson,
arXiv:hep-ph/0505002.
\bibitem{CS} F.E.Close and E.S.Swanson, arXiv:hep-ph/0505206
\bibitem{Eichten} E. Eichten, K. Gottfried, T. Kinoshita, K. D. Lane, and
T. M. Yan, Phys. Rev. D17, 3090 (1978); ibid. D21, 203 (1980).
\bibitem{ABarnes}E. S. Ackleh, T. Barnes, E. S. Swanson, Phys.Rev. D54
6811 (1996) [arXiv:hep-ph/9604355].
\bibitem{Simonov} Yu. A. Simonov, Phys. Atom. Nucl. 66, 2033 (2003)
[ Yad. Fiz. 66 2083 (2003)] [arXiv:hep-ph/0211410].
\bibitem{nEichten} E. J. Eichten, K. Lane, C. Quigg,
Phys. Rev. D69  094019 (2004) [arXiv:hep-ph/0401210].
\bibitem{Barnes0} T. Barnes, arXiv:hep-ph/0412057.
\bibitem{port} E. van Beveren, C. Dullemond, and G. Rupp, Phys. Rev. D21,
772 (1980); G. Rupp, E. van Beveren, and
M.D. Scadron, Phys. Rev. D65, 078501 (2002) [arXiv: hep-ph/0104087]; E.
van Beveren and G. Rupp, arXiv:hep-ph/0304105.
\bibitem{Tornqvist} N. A. Tornqvist, Z. Phys. C68, 674
(1995) [arXiv:hep-ph/9504372]; N. A. Tornqvist and M. Roos, Phys. Rev.
Lett. 76, 1575 (1996) [arXiv: hep-ph/9511210].
\bibitem{Markushin} M. P. Locher, V. E.
Markushin, and H. Q. Zheng, Eur. Phys. J. C4, 317 (1998) [arXiv:
hep-ph/9705230]; V. E.
Markushin, Eur. Phys. J. A8, 389 (2000) [arXiv: hep-ph/0005164].
\bibitem{Pennington} M. Boglione, M.R. Pennington,
Phys.Rev. D65, 114010 (2002) [arXiv: hep-ph/0203149].
\bibitem{Ds} T. Barnes, F. E. Close and H. J. Lipkin, Phys.
Rev. D68, 054006 (2003) [arXiv:hep-ph/0305025]; E. van Beveren and G.
Rupp, Phys. Rev. Lett. 91, 012003 (2003) [arXiv:hep-ph/0305035].
\bibitem{ST} Yu. A. Simonov, J. A. Tjon, Phys. Rev. D70, 114013
(2004) [arXiv:hep-ph/0409361].
\bibitem{Ker}B. O. Kerbikov, Theor. Math. Phys. 65, (1985) 
[Teor. Mat. Fiz. 65, 379 (1985)].
\bibitem{SWein} S. Weinberg, Phys. Rev.
130, 776 (1963); 131, 440 (1963); 137 B672 (1965).
\bibitem{bhm} L.N. Bogdanova, G.M. Hale, and V.E. Markushin, Phys. Rev. C
44, 1289 (1991).
\bibitem{evidence}  V. Baru, J. Haidenbauer, C. Hanhart, Yu. Kalashnikova,
A. Kudryavtsev, Phys. Lett. B586, 53 (2004) [arXiv:hep-ph/0308129].
\bibitem{Flatte} V. Baru, J. Haidenbauer, C. Hanhart, A. Kudryavtsev,
Ulf-G. Meissner, Eur. Phys. J. A23, 523 (2005) [arXiv:nucl-th/0410099].
\bibitem{Alla} A. M. Badalian, V. L. Morgunov, B. L. G.
Bakker, Phys. Atom. Nucl.
63, 1635 (2000) 1635-1639 [Yad. Fiz. 63, 1722 (2000)
[arXiv:hep-ph/9906247].
\bibitem{PDG} S. Eidelman {\it et al} (Particle Data Group), Phys. Lett.
B592, 1 (2004)
\bibitem{Rosner}J. L. Rosner, Phys. Rev. D64, 094002 (2001)
[arXiv:hep-ph/0105327].
\bibitem{Voloshincc}M. B. Voloshin, arXiv:hep-ph/0504197.
\bibitem{Voloshinspin} M. B. Voloshin, Phys. Lett. B604, 69 (2004)
[arXiv:hep-ph/0408321].
\bibitem{SwansonX} E. S. Swanson, Phys. Lett. B588, 189 (2004)
[arXiv:hep-ph/0311229].
\bibitem{Voloshindecay} M. B. Voloshin, Phys. Lett. B579, 316 (2004)
[arXiv:hep-ph/0309307].
\bibitem{Braaten1}E. Braaten, M. Kusunoki, Phys. Rev. D69, 074005 (2004)
[arXiv:hep-ph/0311147].
\bibitem{Braaten2}E. Braaten, M. Kusunoki, S. Nussinov, Phys. Rev. Lett.
93 162001 (2004) [arXiv:hep-ph//0404161].
\bibitem{Braaten3} E. Braaten, M. Kusunoki, arXiv:hep-ph/0506087
\bibitem{BelleY}S.-K. Choi, S. L. Olsen, {\it et al}. [Belle
Collaboration], Phys. Rev. Lett. 94, 182002 (2005) [arXiv:hep-ex/0408126].
\end{thebibliography}
\end{document}